\documentclass[11pt]{article}             
\usepackage{osid2}                         %

\title{Tests of Complete Positivity in Fiber Optics}
\author{Fabio Benatti \\{\footnotesize\it Dipartimento di Fisica Teorica, 
Universit\`a di Trieste \& Istituto Nazionale di Fisica Nucleare, 
Sezione di Trieste, Trieste, Italy}\\[2ex]
Roberto Floreanini\\{\footnotesize\it Istituto Nazionale di Fisica Nucleare, 
Sezione di Trieste, Trieste, Italy}          }

\begin{document}

\maketitle
\begin{abstract}
We consider the propagation of polarized photons in optical
fibers under the action of randomly generated noise.
In such situation, the change in time of the photon polarization 
can be described by a quantum dynamical semigroup. We show that the hierarchy among 
the decay constants of the polarization density matrix elements as prescribed by
complete positivity can be experimentally probed using standard
laboratory set-ups.
\end{abstract}

\section{Introduction}

A quantum system immersed in a large environment represents
the paradigm of a physical situation that is often encountered
in quantum optics, quantum chemistry and atomic physics.
Since it is not isolated, such system is called open: it
exchanges energy and entropy with the external environment, 
whose dynamics is typically unaffected by the presence of 
the small subsystem. The evolution of such open quantum systems,
obtained by tracing over the (infinite) degrees of freedom
of the environment, is in general very complicated,
exhibiting non linearities and memory effects.

However, when the coupling between the two systems is weak,
a very common situation in many practical applications,
the reduced time evolution can be described by a one-parameter
(=time) family of linear maps $\{\gamma_t\}$, a so-called quantum dynamical
semigroup \cite{01}-\cite{05}. These linear transformations map states
of the subsystem, described by density matrices, into
states, while possessing many desirable 
physical properties. In particular, 
these maps exhibit irreversibility, encoded in the
semigroup property ({\it i.e.} in the forward in time 
composition law), decoherence and dissipation,
the typical noisy effects induced by the environment.
Further, these generalized time evolutions turn out 
to be completely positive, a property that guarantees
the physical consistency of the dynamics in all
physical situations.

Nevertheless, in many phenomenological 
treatments of the dynamics of open quantum systems,
this last property is often dismissed as irrelevant,
nothing more than a mathematical artifact \cite{06}-\cite{10}.
The argument supporting this point of view is based 
on the following considerations.
In order to represent a physical
state of the subsystem $S$, a density matrix $\rho$ must be
a positive operator, since its eigenvalues have the
meaning of probabilities;
this is at the root of the statistical interpretation of quantum mechanics.
The time evolution $\rho(0)\mapsto \rho(t)=\gamma_t[\rho(0)]$
must then preserve this fundamental property, and therefore
map a positive initial $\rho(0)$ into a positive final $\rho(t)$.
Such a property of the linear transformation $\gamma_t$ is called
{\it positivity}; it is apparently sufficient to assure the physical 
consistency of the dynamics.

On the other hand, {\it complete positivity}
is a more restrictive requirement than simple positivity: it guarantees
the positivity not only of $\gamma_t$ but also of the dynamics
of a larger system built with
the system $S$ statistically coupled to a second ancillary system
$S'$, which however remains inert under the time evolution. 
The dynamics of this enlarged system is then described
by ${\mit\Gamma}_t=\gamma_t\otimes {\rm id}$, where ${\rm id}$
represents the identity operation. Positivity of ${\mit\Gamma}_t$
for any ancillary system $S'$ means complete positivity
of the map $\gamma_t$. It is worth noting that this property
is intimately related to entanglement,
{\it i.e.} to the possibility that $S$ may have interacted in the
past with $S'$ and become quantum correlated with it \cite{04,11}.

Admittedly, although logically stringent, the above motivation of
why an open system dynamics should be completely positive is
not very appealing from the physical point of view; indeed,
the ancilla system $S'$ is remote from $S$ and fixed.
Nevertheless, a more concrete scenario can be offered,
that of two equal, mutually non interacting systems
$S$, immersed in the same external bath; such a situation
is not uncommon and it is for instance encountered 
in high energy particle physics in
the study of correlated neutral mesons \cite{12,13,14}.
In this case the two systems evolve with the product map
${\mit\Gamma}_t=\gamma_t\otimes\gamma_t$, and one can 
show that the positivity of ${\mit\Gamma}_t$ is equivalent
to the complete positivity of $\gamma_t$ \cite{15}.

Another important advantage of having a completely positive
open system dynamics is that this property fully characterizes
the form of the map $\gamma_t$ \cite{16,17}. In particular,
when $S$ is a two-level systems, as discussed below, 
the decay times of the diagonal $(T_1)$ and
off-diagonal $(T_2)$ elements of the corresponding 
density matrix are seen to satisfy a characteristic
order relation: $2\, T_1\geq T_2$ \cite{01,18}. 
It is precisely
the presence of such hierarchy that is often
questioned in the phenomenological literature
on open quantum systems. 

The aim of the present investigation is to analyze
in detail the possibility of a direct test of
the above inequality in  the laboratory,
using polarized photons. The basic idea is to
study the change in their polarization while the
photons travel across a weakly coupled, noisy environment.
A possible practical realization of this scenario
involve photons with a given polarization injected
inside a high quality, polarization preserving optical fiber,
which is subjected to random noise generated
by external high frequency sound waves; as we shall see,
the random noise will play a role similar of that
of an external environment.
By making the photons traverse the
fiber twice with the insertion of a Faraday mirror at one end, it is possible
to measure the ratio between the diagonal and
off-diagonal elements of the $2\times 2$ density matrix
representing the final photon polarization state, 
and thus determine the combination
$2\, T_1 - T_2$. 

All the necessary techniques needed for the
realization of the apparatus and the realization
of the measure are standard and available
in modern quantum optical laboratories. We hope that our
results will stimulate the actual realization
of the experiment, thus clarifying in a controlled setting
the role of complete positivity in open quantum dynamics.

\section{Master equation}

In describing the evolution of polarized photons in a optical fiber
we shall adopt the standard effective description in terms 
of a two-dimensional Hilbert space,
the space of helicity states \cite{19}-\cite{23}.
Any vector in this space
represents a given polarization and can be identified by two angles
$\theta$ and $\varphi$:
\begin{equation}
|\theta,\varphi\rangle=\cos\theta\, |+\rangle\,
+\, e^{i\varphi}\sin\theta\, |-\rangle\ ,
\label{1}
\end{equation}
where $|+\rangle$ and $|-\rangle$ are two orthonormal basis vectors,
representing linearly polarized states. Another convenient basis in
this space is given by the circularly polarized states:
\begin{equation}
|R\rangle={1\over\sqrt2}\Big(|+\rangle\, +\, i |-\rangle\Big)\ ,\qquad
|L\rangle={1\over\sqrt2}\Big(|+\rangle\, -\, i |-\rangle\Big)\ .
\label{2}
\end{equation}
With respect to this basis, any (partially) polarized photon state can be
represented by a $2\times2$ density matrix $\Sigma$; as already pointed out before,
this is a hermitian, positive operator, {\it i.e.} with positive eigenvalues, 
normalized to have unit trace.

The time evolution of the photons inside the fiber, while
subjected to random noise, can be cast
in a standard Liouville - von Neumann form:
\begin{equation}
{\partial \Sigma(t)\over \partial t}= -i\big[ H_0, \Sigma(t)\big]+ L_t\big[\Sigma(t)\big]\ .
\label{3}
\end{equation}
The first piece on the r.h.s. describes the propagation of the photons
in absence of noise; in the chosen basis, the effective hamiltonian $H_0$
can be cast in the general form:
\begin{equation}
H_0={\omega_0\over2}\, \vec n\cdot\vec\sigma\ ,
\label{4}
\end{equation}
where $\vec\sigma=(\sigma_1,\sigma_2,\sigma_3)$
is the vector of Pauli matrices. For sake of generality,
we have kept $\omega_0$ nonvanishing, in order to take 
into account possible birefringence effects 
due to the fiber mechanical bending.

The fiber is further subjected to random
noise, generated by sound waves of frequency much higher than
the inverse flight time of the photons in the fiber.
As a consequence, the travelling photon will 
propagate in wave guide that is randomly changing; it is as
if it was effectively moving into a rapidly fluctuating medium, 
which can be described by classical stochastic fields.
The action of this fields on the travelling photons
can then be expressed via the commutator with a
time-dependent hermitian matrix $F(t)$,
\begin{equation}
L_t\big[\Sigma(t)\big]=-i\big[ F(t), \Sigma(t) \big]\ ,\qquad F(t)=\vec F(t)\cdot\vec\sigma\ ,
\label{5}
\end{equation}
whose components $F_1(t)$, $F_2(t)$, $F_3(t)$ form a real, stationary Gaussian
stochastic field $\vec F(t)$; in general, they have nonzero 
constant means and translationally invariant correlations:
\begin{equation}
G_{ij}(t-s)\equiv\langle F_i(t)\, F_j(s)\rangle
-\langle F_i(t)\rangle\, \langle F_j(s)\rangle\ ,\quad i,j=1,2,3\ .
\label{6}
\end{equation}

Since the generalized hamiltonian $F(t)$ in (\ref{5}) involves stochastic 
variables,
the density matrix $\Sigma(t)$, solution of the equation of
motion (\ref{3}), is also stochastic. Instead, we are interested in the behaviour
of the reduced density matrix $\rho(t)\equiv\langle \Sigma(t)\rangle$
which is obtained by averaging over the noise; it is $\rho(t)$ that
describes the effective evolution of the photons in the randomly
behaving fiber.
By making the additional assumption that photons
and noise be decoupled at $t=\,0$, so that the initial state is
$\rho(0)\equiv\langle \Sigma(0)\rangle=\Sigma(0)$, a condition very well satisfied 
in practice, an effective master equation for $\rho(t)$
can be derived by going to the interaction representation,
where we set:
\begin{equation}
\widetilde{\Sigma}(t)=e^{it\, H_0}\ \Sigma(t)\ e^{-it\, H_0}\ ,\quad
\vec\sigma(t)=e^{it\, H_0}\ \vec\sigma\ e^{-it\, H_0}\ ,\quad
\widetilde{L}_t[\ \ ]\equiv-i\Big[\vec F(t)\cdot\vec\sigma(t),\ \ \Big]\ .
\label{7}
\end{equation}

The time evolution of the reduced density matrix in this representation,
$\tilde{\rho}(t)\equiv\langle\widetilde{\Sigma}(t)\rangle$, can then be expressed
as a series expansion involving multiple correlations
of the operator $\widetilde{L}_t$:
\begin{equation}
\tilde\rho(t)= {\cal N}_t[\tilde\rho(0)]
\equiv\sum_{k=0}^\infty\int_0^t ds_1\int_0^{s_1}ds_2\cdots \int_0^{s_{k-1}}ds_k\
\langle \widetilde{L}_{s_1} \widetilde{L}_{s_2}\cdots \widetilde{L}_{s_k}\rangle
[\tilde\rho(0)\,]\ .
\label{8}
\end{equation}
The sum ${\cal N}_t[\tilde\rho]\equiv\sum_{k=0}^\infty N_k[\tilde\rho]$ 
in (\ref{8}) with $N_0[\tilde\rho]\equiv\tilde\rho$ 
can be formally inverted, and a suitable resummation
gives (a dot represents time derivative) \cite{24}:
\begin{equation}
{\partial \tilde{\rho}(t)\over\partial t}=\dot{\cal N}_t\, {\cal N}_t^{-1}[\tilde\rho(t)]=
\Big\{\dot{N}_1+\big(\dot{N}_2-\dot{N}_1\, N_1\big)+\ldots\Big\}[\tilde\rho(t)]\ .
\label{9}
\end{equation}

Since the interaction of the travelling photons with the stochastic 
field is weak,
one can focus the attention on the dominant terms of the previous expansion, 
neglecting all contributions higher than the second-order ones.
Further, since the characteristic fluctuation time of the random noise
is by assumption much smaller than the typical photon time of flight
across the fiber,
the memory effects implicit in (\ref{9}) should not be physically relevant
and the use of the Markovian approximation justified. This is implemented 
in practice by extending to infinity the upper limit of the integrals
appearing in $\dot{N}^{(2)}$ and $N^{(1)}$ \cite{01}-\cite{03}.

By returning to the Schr\"odinger representation, one finally obtains \cite{04,25}
\begin{equation}
{\partial\rho(t)\over\partial t}=-i\big[ H,\, \rho(t)\big]
+L\big[\rho(t)\big]\ ,
\label{10}
\end{equation}
where
\begin{equation}
H=H_0 + H_1 + H_2\equiv\vec\omega\cdot\vec\sigma\ ,
\label{11}
\end{equation}
and
\begin{equation}
L[\rho]={1\over2}\sum_{i,j=1}^3 {\cal C}_{ij}
\Big[2\sigma_j \rho\, \sigma_i
- \{\sigma_i\sigma_j\, ,\, \rho\}\Big]\ .
\label{12}
\end{equation}

The effective hamiltonian $H$ differs from the one in absence of noise
$H_0$ by first order terms (coming from the piece $\dot{N}^{(1)}$
in (\ref{9})) depending on the noise mean values:
\begin{equation}
H_1= \langle \vec F(t)\rangle\cdot \vec\sigma\ ,
\label{13}
\end{equation}
and by second-order contributions (coming from the second-order terms 
in (\ref{9})),
\begin{equation}
H_2=\sum_{i,j,k=1}^3\,\epsilon_{ijk}\ C_{ij}\, \sigma_k\ ,
\label{14}
\end{equation}
involving the noise correlations (\ref{6}) through the 
time-independent combinations:
\begin{equation}
C_{ij}=\sum_{k=1}^3\int_0^\infty dt\ G_{ik}(t)\ U_{kj}(-t)\ ,
\label{15}
\end{equation}
where the $3\times3$ orthogonal matrix $U(t)$ is defined by the following
transformation rule:
$e^{it\, H_0}\ \sigma_i\ e^{-it\, H_0}=\sum_{j=1}^3 U_{ij}(t)\ \sigma_j$.
On the other hand, the contribution $L[\rho\,]$ in (\ref{12})
is a time-independent, trace-preserving linear map
involving the symmetric coefficient matrix ${\cal C}_{ij}\equiv C_{ij}+C_{ji}$,
which needs to be positive, ${\cal C}_{ij}\geq0$, in order to generate
a completely positive dynamics \cite{01}-\cite{04}.
It introduces irreversibility, inducing
in general dissipation and loss of quantum coherence.

Altogether, equation (\ref{10})-(\ref{12}) generates a semigroup of linear maps,
${\mit\Gamma}_t:\ \rho(0)\mapsto\rho(t)\equiv{\mit\Gamma}_t[\rho(0)]$,
for which composition is defined only forward in time:
${\mit\Gamma}_t\circ{\mit\Gamma}_s={\mit\Gamma}_{t+s}$, with $t,s\geq0$;
it is usually referred to as a quantum dynamical semigroup \cite{01}-\cite{04}.

It is convenient to expand the photon
density matrix in terms of the Pauli matrices and the identity $\sigma_0$,
\begin{equation}
\rho={1\over2}\Big[\sigma_0+\vec\rho\cdot\vec\sigma\Big]\ .
\label{16}
\end{equation}
Then, the linear equation (\ref{10}) reduces to a diffusion equation
for the components $\rho_1$, $\rho_2$, $\rho_3$ of the (Bloch) vector
$\vec\rho$:
\begin{equation}
{\partial \vec\rho(t)\over\partial t}=-2{\cal H}\, \vec\rho(t)\ ;
\label{17}
\end{equation}
the entries of the $3\times3$ matrix $\cal H$ can be expressed
in terms of the coefficients $\omega_i$ and ${\cal C}_{ij}$
appearing in the hamiltonian and noise contribution 
in (\ref{11}), (\ref{12}) \cite{01,04}:
\begin{equation}
{\cal H}=\left[
\matrix{a&b+\omega_3&c-\omega_2\cr     
b-\omega_3&\alpha&\beta+\omega_1\cr                                     
c+\omega_2&\beta-\omega_1&\gamma\cr}
\right]\ ,
\label{18}
\end{equation}
with $a={\cal C}_{22}+{\cal C}_{33}$, $\alpha={\cal C}_{11}+{\cal C}_{33}$,
$\gamma={\cal C}_{11}+{\cal C}_{22}$, $b=-{\cal C}_{12}$,
$c=-{\cal C}_{13}$, $\beta=-{\cal C}_{23}$. 
The condition of complete positivity, ${\cal C}_{ij}\geq0$,
can then be expressed more explicitly through the following inequalities:
\begin{eqnarray}
\nonumber
&&2R\equiv\alpha+\gamma-a\geq0\ ,\qquad RS\geq b^2\ ,\\
\nonumber
&&2S\equiv a+\gamma-\alpha\geq0\ ,\qquad RT\geq c^2\ ,\\
\nonumber
&&2T\equiv a+\alpha-\gamma\geq0\ , \qquad ST\geq\beta^2\ ,\\
\label{19}
&&RST\geq 2\, bc\beta+R\beta^2+S c^2+T b^2\ .\\
\nonumber
\end{eqnarray}
\vskip -.8cm
The solution of (\ref{17}) involves the exponentiation of the matrix $\cal H$,
\begin{equation}
\vec\rho(t)={\cal M}(t)\ \vec\rho(0)\ ,\qquad {\cal M}(t)=e^{-2{\cal H} t}\ .
\label{20}
\end{equation}
From this relation, one immediately sees that the $3\times 3$ matrix ${\cal M}(t)$
represents the (reduced) Mueller matrix connecting the initial
Stokes vector $\vec\rho(0)$ with the evolved one $\vec\rho(t)$ 
at time $t$ \cite{21,22,23,26}.

\section{Complete positivity}

As shown in the previous section, the entries of the matrix
$\cal H$ in (\ref{18}) are directly related to the
stochastic field correlations in (\ref{6}). For a generic noise,
one expects all parameters in $\cal H$ to be nonvanishing; only
physical considerations may allow a simplification and therefore
a manageable, explicit expression for $\vec\rho(t)$.

For a typical random noise, one can generically assume the
correlation functions to have an exponentially damped form. Further,
the off-diagonal correlations are usually much suppressed
with respect to the diagonal ones, so that, without
much loss of generality, one can write:
\begin{equation}
G_{ij}(t)=G_i\ e^{-\lambda_i\, |t|}\ \delta_{ij}\ ,\qquad G_i,\ \lambda_i\geq0\ .
\label{21}
\end{equation}
Using the expression (\ref{4}) for the hamiltonian $H_0$, 
one easily finds the form of the matrix $U_{ij}(t)$ giving
the time evolution of the Pauli matrices,
\begin{equation}
U_{ij}(t)=n_i\, n_j +(\delta_{ij}-n_i\, n_j)\ \cos\omega_0 t
-\sum_{k=1}^3\varepsilon_{ijk} n_k\ \sin\omega_0 t
\label{22}
\end{equation}
and then through the definition (\ref{15}), one finally obtains:
\begin{equation}
C_{ij}=\lambda_i\Lambda_i\bigg[\delta_{ij}+
{\omega_0^2\over\lambda_i^2}\, n_i\, n_j+
{\omega_0\over\lambda_i}\sum_{k=1}^3\varepsilon_{ijk} n_k\bigg]\ ,
\label{23}
\end{equation}
where
\begin{equation}
\Lambda_i={G_i\over \lambda_i^2+\omega_0^2}\ .
\label{24}
\end{equation}
The symmetric and antisymmetric pieces of $C_{ij}$ give the noise contributions
to the dissipative (\ref{12}) and hamiltonian (\ref{14}) parts of the evolution
equation (\ref{10}).

In order to further simplify the treatment, we shall assume the unit
vector $\vec n$ defining the starting hamiltonian $H_0$ to be directed
along the third axis and the stochastic field $\vec F(t)$ to have zero mean.
In this case, the effective hamiltonian in
(\ref{11}) remains proportional to $\sigma_3$,
$H=\omega\sigma_3$, with a new frequency $\omega\equiv\omega_3$
explicitly containing zero and second order
pieces: $\omega=\omega_0/2 +\omega_0\, (\Lambda_1+\Lambda_2)$.
Similarly, also the dissipative contributions simplifies,
since one finds $c=\beta=\,0$, while
\begin{eqnarray}
\nonumber
&&a=2\lambda_2\Lambda_2+2 {G_3\over\lambda_3}\ ,\qquad
b=\omega_0\big(\Lambda_2 - \Lambda_1\big)\ ,\\
\label{25}
&&\alpha=2\lambda_1\Lambda_1+2{G_3\over\lambda_3}\ ,
\qquad \gamma=2\lambda_1\Lambda_1+2\lambda_2\Lambda_2\ .\\
\nonumber
\end{eqnarray}
Notice that the conditions assuring the complete positivity 
of the time evolution given by the
inequalities in (\ref{19}) are not all automatically satisfied
by the assignments in (\ref{25}). The following relation
needs to be imposed, 
$\omega_0^2(\Lambda_2-\Lambda_1)^2\leq 4\,\lambda_1\lambda_2\Lambda_1\Lambda_2$:
it can always be fulfilled by a suitable choice of the noise parameters
in (\ref{21}). 

This situation is not exceptional: the derivation of a physically
consistent, Markovian reduced dynamics
starting from the exact Liouville - von Neumann equation in (\ref{3})
is in general very involved and can be treated with the necessary
mathematical rigor only in special cases \cite{01}-\cite{03}.
In general, as briefly discussed in Section 2,
one instead resorts to various approximations,
justified by physical considerations.
However, this naive procedure does not always lead
to physically consistent reduced dynamics \cite{04}: further conditions,
like that of complete positivity, need to be imposed at the end.
This explains why this property is often dismissed as irrelevant in
phenomenological applications, and therefore why it is so important
to verify its fulfillment in an experimentally controlled setting.

With the above assignments, the exponentiation of $\cal H$
can be easily evaluated in closed form, and for the Mueller
matrix one explicitly finds:
\begin{equation}
{\cal M}(t)=
\left[
\matrix{e^{-(a+\alpha)\, t}\ A_+(t) &  e^{-(a+\alpha)\, t}\ B_+(t) & 0 \cr     
e^{-(a+\alpha)\, t}\ B_-(t) &  e^{-(a+\alpha)\, t}\ A_-(t) & 0 \cr                                   
0 & 0 &e^{-2\gamma\, t}\cr}
\right]\ ,
\label{26}
\end{equation}
where
\begin{eqnarray}
\nonumber
&&A_\pm(t)=\cos (2\Omega\,t) \pm {\alpha-a\over2\Omega}\, \sin (2\Omega\, t)\ ,\\
\label{27}
&&B_\pm(t)=-{b\pm\omega\over\Omega}\sin (2\Omega\, t)\ ,\\
\nonumber
\end{eqnarray}
with $\Omega=\big[\omega^2 -b^2 -(a-\alpha)^2/4\big]^{1/2}$. The evolution matrix
in (\ref{26}) contains both oscillating and damping terms, and can be considered
as a generalization of a rotator \cite{21}-\cite{23}.
Notice, however, that the oscillator behaviour
depends on the magnitude of effective frequency $\omega$ with respect
to the dissipative parameters $a$, $b$ and $\alpha$; when 
$\omega < \big[b^2 +(a-\alpha)^2/4\big]^{1/2}$, the frequency $\Omega$
becomes purely imaginary and ${\cal M}(t)$ contains only exponential terms.

In any case, for large times, {\it i.e.} for infinitely long fibers, 
the dumping terms always dominate \cite{27} and
the Mueller matrix becomes that of a total depolarizer;
in other words, as result
of the action of the stochastic noise, the photon density matrix $\rho(t)$
will become asymptotically proportional to the unit matrix
({\it i.e.} \hbox{$\vec\rho(t)\sim 0$}), independently
from the initial state $\rho(0)$. 

Through the definition (\ref{16}), 
the evolution matrix (\ref{26}) gives the
time behaviour of the entries of the photon
density matrix $\rho(t)$, which in turn can be experimentally
determined using suitable tomographic procedures.
Therefore, at least in principle, thanks to the different 
time-dependence of the entries in (\ref{26}),
one can measure the magnitude
of all the dissipative parameters $a$, $b$, $\alpha$ and $\gamma$
and thus check the conditions of complete positivity,
that in particular requires: $a+\alpha-\gamma\geq0$.

Unfortunately, this can not be done by a single passage of the
photons in the noisy fiber, since in this case the time $t$ appearing
in (\ref{26}) is fixed, being the time of flight along the fiber.
In order to isolate the exponential damping terms from the
oscillating ones, $t$ needs to be varied. This can be achieved
by placing at the end of the fiber a Faraday mirror \cite{21},
which inverts the polarization of the photons, while reflecting them
back into the fiber. In this way, travelling backward, 
the rotation induced by the standard hamiltonian $H_0$ 
on the photon polarization is ``undone'',
while the dissipative effects due to the stochastic noise 
further accumulates. Indeed, the backward evolution
is given by a Mueller matrix $\widetilde{\cal M}(t)$
still of the form (\ref{26}),
but with the frequency $\omega_0$ replaced by $-\omega_0$,
which in turn gives $\omega\to -\omega$ and $b\to -b$,
while leaving the remaining dissipative parameters unchanged.

Thanks to the semigroup property of the dynamics,
the complete evolution for the double passage of the photons in the fiber
is obtained by composing the two Mueller matrices,
and therefore, after having travelled for a time $2t$ inside
the noisy fiber, the Stokes parameters representing the photon
polarization state become at the end:
\begin{equation}
\vec\rho(2t)=\widetilde{\cal M}(t)\cdot {\cal M}(t)\, \vec\rho(0)\ .
\label{28}
\end{equation}

Let us now consider an initially linearly polarized photon,
so that $\vec\rho(0)\equiv\vec\rho^{\, (+)}(0)=(1, 0, 0)$.
After having passed once trough the noisy fiber, this Stokes
vector becomes $\vec\rho^{\, (+)}(t)$, as in (\ref{20}), while
further evolves to $\vec\rho^{\, (+)}(2t)$, as in (\ref{28}), after having
travelled backwards to the beginning of the fiber. 
Similar results hold for a circularly polarized initial
photon, for which: $\vec\rho(0)\equiv\vec\rho^{\, (R)}(0)=(0, 0, 1)$.
One can then form the following combination of components
of the above Stokes vectors:
\begin{equation}
{\cal R}(t)={1\over \rho^{(R)}_3(t)}\Bigg[
\rho^{(+)}_1(2t) +\rho^{(+)}_2(2t)\ {\rho^{(+)}_1(t)\over \rho^{(+)}_2(t)}\Bigg]\ .
\label{29}
\end{equation}
Using the explicit expressions for these Stokes components as obtained
from (\ref{26}), one easily finds: ${\cal R}(t)=\exp[{-2(a+\alpha-\gamma)\,t}]$.
Recall that complete positivity requires the combination $a+\alpha-\gamma$
to be nonnegative. Therefore, by measuring the components of the
Stokes vectors appearing in (\ref{29}), one can determine the 
quantity ${\cal R}(t)$ and thus check whether or not%
\footnote{Note that this inequality determines the sign of the
combination $a+\alpha-\gamma$ for any $t$, {\it i.e.} irrespective
of the length of the fiber, provided it behaves in a stochastic way.
This fact might help reduce the systematic uncertainties in the
actual experimental test of (\ref{30}).}
\begin{equation}
{\cal R}(t)\leq 1\ .
\label{30}
\end{equation}
Although this condition is in general only necessary for complete positivity,
it becomes also sufficient for the phenomenologically relevant
situation for which $\Lambda_1=\Lambda_2$. Recalling (\ref{25}), in this case
one has $b=\,0$ and further $a=\alpha$, so that $2\alpha\geq\gamma$
is the only surviving inequality of those listed in (\ref{19}).

The two nonvanishing dissipative parameters $\alpha$ and $\gamma$
have now the meaning of inverse relaxation times for the
off-diagonal and diagonal entries of the photon
density matrix $\rho(t)$ \cite{01}. They are usually called
$1/T_2$ and $1/T_1$, respectively. The condition
that assures the complete positivity of the open dynamics
for photons travelling along the noisy fiber is therefore
precisely $2 T_1\geq T_2$, as was mentioned in the
introductory remarks. 

In summary, we have shown that the condition assuring the
complete positivity of the dynamics of polarized photons 
in a noisy fiber can be cast in the form (\ref{30}).
By measuring the combination
$\cal R$ in (\ref{29}), it can be tested using
set-ups and techniques that are routinely
used in optical laboratories.
We find this possibility very intriguing and hope
will trigger the interest of the vast community
of specialists working in quantum optics experiments.

\end{document}